\documentclass[11pt]{article}

\usepackage{mathrsfs, epsf}

\begin{document}

\title{Stress-energy tensor for a quantised bulk scalar field in the Randall-Sundrum brane model}
\author{A.~Knapman\thanks{e-mail address: {\tt a.h.knapman@ncl.ac.uk}}\ \ and D.~J.~Toms\thanks{e-mail address: {\tt d.j.toms@newcastle.ac.uk}}\\
School of Mathematics and Statistics,\\ University of Newcastle upon Tyne,\\
Newcastle Upon Tyne, England NE1 7RU}
\maketitle

\begin{abstract}
We calculate the vacuum expectation value of the stress-energy tensor for a quantised bulk scalar field in the Randall-Sundrum model, and discuss the consequences of its local behaviour for the self-consistency of the model. We find that, in general, the stress-energy tensor diverges in the vicinity of the branes. Our main conclusion is that the stress-energy tensor is sufficiently complicated that it has implications for the effective potential, or radion stabilisation, methods that have so far been used.
\end{abstract}

\section{Introduction}

It is widely being considered whether our spacetime has more than four dimensions. Higher dimensional models have arisen in the context of the search for a unified field theory since the proposals of Nordstr\"om, Kaluza and Klein \cite{kk} in the early years of general relativity. These ideas have been extended over time, and have recently been revived \cite{antoniadis} in the form of the so-called ``brane-world'' models, in which the (3+1)-dimensional universe we observe exists as a membrane, or 3-brane, in a higher dimensional spacetime. In the framework of string theory, Hora\u va and Witten \cite{horava} have constructed such a model that may accommodate a realistic gauge theory.

In 1999, Randall and Sundrum (RS) \cite{rs} proposed that a scenario closely related to the Hora\u va-Witten model could naturally explain the problem of the large hierarchy between the fundamental scale of quantum gravity, the Planck scale $M_{Pl}\sim 10^{15}$ TeV, and the scale of the standard model of particle physics, the electroweak scale $M_{EW}\sim 1$ TeV. In this scenario, two flat, parallel 3-branes of opposite tension sit at the fixed points of a $S^1/${\bf Z}$_2$ orbifold. The ``bulk'' between the two branes forms a slice of 5-dimensional anti-de Sitter spacetime. The geometry is then non-factorisable, the four-dimensional part of the metric being multiplied by a ``warp'' factor, an exponential function of the fifth dimension:
\begin{equation}
  ds^2 = e^{-2\sigma (y)} \eta _{\mu \nu} dx^\mu dx^\nu - dy^2 , \label{lmetric}
\end{equation}
where $\sigma (y)=-k|y|$ ($k\sim M_{Pl}$) and $-\pi r\le y\le \pi r$ is the extra dimension. The branes are located at $y=0$ and $y=\pi r$. The brane at $y=\pi r$ known as the visible brane and represents the universe as we know it. Conversely, that at $y=0$ is called the hidden brane. In the original RS model, gravity is the only field present in all five dimensions. All of the standard model fields are confined to the visible brane. Alternatives to this somewhat artificial restriction have been investigated by many authors (see e.g.~\cite{goldbergerwiseprd60}). It is the possibility of bulk matter fields, in particular scalar fields, that we deal with in this paper.

The remarkable point about the RS model is that any mass parameter $m_0$ on the visible brane in the fundamental higher dimensional theory will correspond to a physical mass of $e^{-k\pi r}m_0$ when measured in four dimensions. By taking $kr \simeq 12$, it is possible to generate a TeV mass scale from the Planck scale, without introducing any large hierarchies between the fundamental parameters of the theory.

However, in the basic model, the compactification radius $r$ is left undetermined. In order to make the model physically acceptable, a mechanism for stabilising $r$ must be found. A classical mechanism has been suggested in \cite{goldbergerwiseprl}.

In the context of the older Kaluza-Klein theories, Candelas and Weinberg \cite{candelasweinberg} in 1984 found that quantum effects from matter and gravity fields could fix the size of the extra dimensions in a natural way. Various authors have investigated whether a similar procedure could fix the value of $r$ in the RS model \cite{Garriga, toms, GoldbergerRothstein}.

In \cite{flachitoms}, a detailed analysis was given of the requirement of the self-consistency of the model, that the model be a solution of the quantum-corrected Einstein equations, showing that the problem is not just one of minimising the vacuum energy density of the bulk field. The authors found that it is possible to find self-consistent solutions that solve the hierarchy problem only at the price of severely fine-tuning the brane tensions. One of the present authors is currently applying their method to the case of de-Sitter branes, which is of interest to inflationary cosmology \cite{flachietal}.

These, and indeed most, calculations have been performed by introducing a bulk scalar field to the model and reducing the theory to four dimensions by summing or integrating over the extra dimension. However, important effects can be overlooked in this approach. In this paper, we investigate the local behaviour of the full stress-energy tensor for a quantised bulk scalar field, and discuss what the consequences are for the self-consistency of the RS model from this standpoint. We show that the problem of self-consistency and stabilisation of the RS model is more complicated than has generally been assumed. Other related work includes \cite{setare, kanti, mukohyama}.

\section{Set-up and general method}

For ease of calculation, we first rewrite the metric (\ref{lmetric}) in Riemannian form so that all momenta will be positive-definite:
\begin{equation}
  d\hat{s}^2 = \hat{g}_{\hat{\mu } \hat{\nu }} d\hat{x}^{\hat{\mu }} d\hat{x}^{\hat{\nu }} = e^{-2\sigma (y)} \delta _{\mu \nu} dx^\mu dx^\nu + dy^2 , \label{rmetric}
\end{equation}
where a caret denotes a $(D+1)$-dimensional quantity: $\hat{\mu } = 0, 1, ..., D$, so that $\hat{x}^D = y$, whereas $\mu = 0, 1, ..., (D-1)$.

Into the bulk of the above background spacetime, we place a single real scalar field $\phi $ of mass $m$. We allow a general coupling to the scalar curvature $\hat{R}$ with dimensionless constant $\xi $. The action $S$ of the field is
\begin{equation}
  S = \frac{1}{2}\int d^{D+1} \hat{x}\, |\det \hat{g}|^{1/2} [ \hat{g}^{\hat{\mu } \hat{\nu }} \partial _{\hat{\mu}} \phi \partial _{\hat{\nu}} \phi - m^2 \phi ^2 - \xi \hat{R} \phi ^2 ]. \label{action}
\end{equation}
The stress-energy tensor for such a scalar field is given by
\begin{eqnarray}
  \hat{T}_{\hat{\mu } \hat{\nu }} & = & \partial _{\hat{\mu}} \phi \partial _{\hat{\nu}} \phi - \frac{1}{2}\hat{g}_{\hat{\mu } \hat{\nu }} \hat{g}^{\hat{\alpha } \hat{\beta }}\partial _{\hat{\alpha }} \phi \partial _{\hat{\beta }} \phi - \frac{1}{2}\hat{g}_{\hat{\mu } \hat{\nu }} m^2 \phi ^2 + \xi \hat{G}_{\hat{\mu } \hat{\nu }} \phi ^2 \nonumber \\
  &  & +\, \xi \left[ \hat{g}_{\hat{\mu } \hat{\nu }} \hat{g}^{\hat{\alpha } \hat{\beta }} (\phi ^2)_{;\hat{\alpha } \hat{\beta }} - (\phi ^2)_{;\hat{\mu } \hat{\nu }} \right], \label{T}
\end{eqnarray}
where $\hat{G}_{\hat{\mu } \hat{\nu }}$ is the Einstein tensor, which is given by
\begin{eqnarray}
  \hat{G}_{\hat{\mu } \hat{\nu }} = \frac{D}{2}(D-1)k^2 \hat{g}_{\hat{\mu } \hat{\nu }} \label{einstein}
\end{eqnarray}
away from the branes.

Rather than working directly with with this expression for the stress tensor, we re-express the stress tensor as a vacuum expectation value in terms of the Green function $\hat{G}(\hat{x}, \hat{x}')$ for the field.

Expressed as a $D$-dimensional Fourier transform, the Green function for a scalar field in a slice of $(D+1)$-dimensional anti-de Sitter spacetime is \cite{gherghettapomarol}:
\begin{equation}
  \hat{G}(\hat{x}, \hat{x}') = \int \frac{d^{D}p}{(2\pi)^{D}} e^{ip.(x-x')}G_p(y, y'), \label{greenfn}
\end{equation}
with
\begin{eqnarray}
  G_p(y, y') & = & \frac{\pi}{2k}e^{\frac{D}{2} (\sigma_> +\sigma_<)} \nonumber \\
    &   & \times \left[ \frac{j_{\nu}(ip/ak)H_{\nu}^1 (ipe^{\sigma _>}/k)- h_{\nu}^1 (ip/ak)J_{\nu}(ipe^{\sigma _>}/k)}{j_{\nu}(ip/ak)h_{\nu}^1 (ip/k)- h_{\nu}^1 (ip/ak)j_{\nu}(ip/k)} \right] \nonumber \\
    &   & \times \left[ j_{\nu}(ip/k)H_{\nu}^1 (ipe^{\sigma _<}/k)- h_{\nu}^1 (ip/k)J_{\nu}(ipe^{\sigma _<}/k) \right] ,\label{FTgreenfn}
\end{eqnarray}
where $J_{\nu}$ is a Bessel function and $H_{\nu}^1 $ is the Hankel function of the first kind, both of order $\nu$. We have defined:
\begin{eqnarray}
  j_{\nu}(z) & = & \frac{1}{2}D(1-4\xi)J_{\nu}(z) + z\frac{d}{dz}J_{\nu}(z), \nonumber \\
  h_{\nu}^1 (z) & = & \frac{1}{2}D(1-4\xi)H_{\nu}^1 (z) + z\frac{d}{dz}H_{\nu}^1 (z), \nonumber \\
  \nu & = & \sqrt{\frac{m^2}{k^2}+\frac{D^2}{4}-\xi D(D+1)}, \label{jandh}
\end{eqnarray}
and have written $a = e^{-k\pi r}$. We define
\begin{equation}
  \sigma _> = \left\{ \begin{array}{ll}
                       \sigma (y) & \mbox{for $y>y'$} \\
		       \sigma (y') & \mbox{for $y<y'$}
		     \end{array}
              \right.
\end{equation}
and
\begin{equation}
  \sigma _< = \left\{ \begin{array}{ll}
                       \sigma (y') & \mbox{for $y>y'$} \\
		       \sigma (y) & \mbox{for $y<y'$}
		     \end{array}
              \right.
\end{equation}
Now consider
\begin{equation}
  \hat{G}(\hat{x}, \hat{x}') = \langle 0|\phi (\hat{x})\phi (\hat{x}')|0\rangle . \label{G}
\end{equation}
Using this, together with (\ref{greenfn}), it can be shown that
\begin{eqnarray}
  \langle \phi ^2 \rangle & = & \hat{G}(\hat{x}, \hat{x}')|_{\hat{x}'=\hat{x}}, \nonumber \\
  \langle \partial _{\mu } \phi \partial _{\nu } \phi \rangle & = & - \partial _{\mu } \partial _{\nu } \hat{G}(\hat{x}, \hat{x}')|_{\hat{x}'=\hat{x}}, \nonumber \\
  \langle \partial _y \phi \partial _y \phi \rangle & = & \partial _y \partial _{y'} \hat{G}(\hat{x}, \hat{x}')|_{\hat{x}'=\hat{x}}, \nonumber \\
  \langle \partial _{\mu } \phi \partial _y \phi \rangle & = & 0, \nonumber \\
  \langle (\phi ^2)_{;\mu \nu } \rangle & = & -k e^{-2\sigma }\delta _{\mu \nu } (\partial _y + \partial _{y'})\hat{G}(\hat{x}, \hat{x}')|_{\hat{x}'=\hat{x}}, \nonumber \\
  \langle (\phi ^2)_{;yy} \rangle & = & (2\partial _y \partial _{y'} + \partial _y^2 + \partial _{y'}^2)\hat{G}(\hat{x}, \hat{x}')|_{\hat{x}'=\hat{x}}, \nonumber \\
  \langle (\phi ^2)_{;\mu y} \rangle & = & 0. \label{vevs}
\end{eqnarray}
We also note that
\begin{equation}
  \partial _{\mu } \partial _{\nu } \hat{G}(\hat{x}, \hat{x}')|_{\hat{x}'=\hat{x}} = -\delta _{\mu \nu}\frac{1}{D}\int \frac{d^{D}p}{(2\pi)^{D}} p^2 G_p(y,y')|_{y'=y}.
\end{equation}
The vacuum expectation value of the stress tensor (\ref{T}) can therefore be written in terms of the Green function as
\begin{eqnarray}
  \langle \hat{T}_{\mu \nu}\rangle & = & \delta _{\mu \nu}\int \frac{d^{D}p}{(2\pi)^{D}} \left[ \left( \frac{1}{D}-\frac{1}{2} \right) p^2 + \left( 2\xi-\frac{1}{2} \right) e^{-2\sigma } \partial _y \partial _{y'} \right. \nonumber \\
  &   & + \xi e^{-2\sigma }(\partial _y \partial _y + \partial _y \partial _{y'}) + \xi \frac{D}{2}(D-1)k^2 e^{-2\sigma} \nonumber \\
  &   & \left. -\frac{1}{2}m^2 e^{-2\sigma} -\xi (D-1)ke^{-2\sigma}(\partial _y +\partial _{y'})\right] G_p(y,y')|_{y'=y}, \nonumber \\
  \langle \hat{T}_{yy}\rangle & = & \int \frac{d^{D}p}{(2\pi)^{D}} \left[ \frac{1}{2}\partial _{y}\partial _{y'} - \frac{1}{2}e^{2\sigma } p^2 + \xi \frac{D}{2}(D-1)k^2 \right. \nonumber \\ 
  &   & \left. -\frac{1}{2}m^2 -\xi Dk(\partial _{y} +\partial _{y'}) \right] G_p(y,y')|_{y'=y}, \label{genT}
\end{eqnarray}
where we have used equation (\ref{einstein}) for the Einstein tensor. The off-diagonal elements are zero.

Substituting $G_p$ from (\ref{FTgreenfn}) into the above expression gives the stress tensor for a bulk scalar field in the Randall-Sundrum model before renormalisation. Care must be taken in taking the derivatives of $G_p (y,y')$. We use the symmetrised relation
\begin{equation}
  \mathscr{D} G_p (y,y')|_{y'=y} = \lim _{y' \rightarrow y} \frac{1}{2} \mathscr{D} (G_p (y,y')|_{y'<y} + G_p (y,y')|_{y'>y}), \label{diffrel}
\end{equation}
where $\mathscr{D}$ is a differential operator.

The expressions for the stress tensor components so obtained are very complicated in general. To perform the integrals over $p$ and to carry out the necessary renormalisations, numerical methods must be used in all but the simplest case of massless, conformally coupled fields.

\section{Massless, conformally coupled fields}

The case of massless, conformally coupled fields is especially simple. The Bessel functions are of order $\nu = 1/2$, so that they become elementary trigonometric functions, allowing the stress tensor to be calculated exactly. This provides a useful check on the method.

For the conformally coupled case, $m = 0$ and $\xi =(D-1)/4D$, so that we obtain:
\begin{equation}
  G_{p}(y,y') = \frac{e^{\frac{(D-1)}{2}(\sigma_> +\sigma_<)}}{p\sinh \frac{p}{k}(e^{k\pi r}-1)} \cosh \frac{p}{k}(e^{\sigma _{<}}-1) \cosh \frac{p}{k}(e^{\sigma _{>}}-e^{k\pi r}), \label{ccGp}
\end{equation}
for the Fourier transform of the Green function. This is the same as the corresponding function in flat space multiplied by $e^{\frac{(D-1)}{2}(\sigma_> +\sigma_<)}$.

The stress tensor (\ref{genT}) takes the form
\begin{eqnarray}
  \langle \hat{T}_{\mu \nu}\rangle & = & \delta _{\mu \nu} \int \frac{d^{D}p}{(2\pi)^{D}} \left[ \left( \frac{1}{D}-\frac{1}{2}\right) p^2 -\frac{1}{2D}e^{-2\sigma }\partial _{y}\partial _{y'} \right. \nonumber \\
  &   & + \frac{D-1}{4D}e^{-2\sigma }(\partial _{y}\partial _{y} + \partial _{y'}\partial _{y'}) + \frac{(D-1)^2 }{8}k^2 e^{-2\sigma } \nonumber \\  
  &   & \left. - \frac{(D-1)^2 }{4D}ke^{-2\sigma }(\partial _{y} +\partial _{y'}) \right] G_{p}(y,y')| _{y'=y}\; , \nonumber \\
  \langle \hat{T}_{yy}\rangle & = & \int \frac{d^{D}p}{(2\pi)^{D}} \left[ \frac{1}{2}\partial _{y}\partial _{y'} - \frac{1}{2}e^{2\sigma } p^2 + \frac{(D-1)^2 }{8}k^2 \right. \nonumber \\ 
  &   & \left. - \frac{(D-1)}{4}k(\partial _{y} +\partial _{y'}) \right] G_{p}(y,y')| _{y'=y}\; .\label{Tcc}
\end{eqnarray}
Substituting (\ref{ccGp}) into (\ref{Tcc}), we find that
\begin{eqnarray}
  \langle \hat{T}_{\mu \nu}\rangle & = & -\delta _{\mu \nu}\frac{1}{2D}e^{(D-1)\sigma }\int \frac{d^{D}p}{(2\pi)^{D}}\, p\, \coth \frac{p}{k} (e^{k \pi r}-1), \nonumber \\
  \langle \hat{T}_{yy}\rangle & = & \frac{1}{2}e^{(D+1)\sigma }\int \frac{d^{D}p}{(2\pi)^{D}}\, p\, \coth \frac{p}{k} (e^{k \pi r}-1).\label{Tcc2}
\end{eqnarray}
The stress tensor is therefore traceless, as it should be in the conformal case, so we only need to consider the $(0,0)$ component. First, we switch to polar coordinates and write
\begin{eqnarray}
  \langle \hat{T}_{00}\rangle & = & -\frac{1}{2D}e^{(D-1)\sigma } \frac {1}{2^{D-1} \pi ^{D/2} \Gamma (D/2)} I \, , \nonumber \\
  I & = & \int _0 ^\infty dp\, p^D \coth \frac{p}{k} (e^{k \pi r}-1). \label{T00cc}
\end{eqnarray}
The integral $I$ requires regularisation, as $\coth \frac{p}{k} (e^{k \pi r}-1) \rightarrow 1$ as $p \rightarrow \infty$.

First, we rewrite the integral as
\begin{equation}
  I = \int _0 ^\infty dp\, p^D \left( 1 + \frac{2e^{-2\beta p/k}}{1-e^{-2\beta p/k}} \right) , \label{Iinf}
\end{equation}
where we have put $\beta = e^{k \pi r}-1$. The first term in the integrand is responsible for the divergence. We employ a $y$-dependent cutoff $e^{-\sigma} \Lambda$, to obtain
\begin{eqnarray}
  I & = & \lim _{\Lambda \rightarrow \infty} \int _0 ^{e^{-\sigma} \Lambda} dp\, p^D + 2\int _0 ^\infty dp\, p^D \left( \frac{2e^{-2\beta p/k}}{1-e^{-2\beta p/k}} \right) \nonumber \\
  & = & \lim _{\Lambda \rightarrow \infty} \frac{e^{-(D+1)\sigma }\Lambda ^{D+1}}{D+1} + 2\left( \frac{k}{2\beta }\right) ^{D+1} \Gamma (D+1)\, \zeta _R (D+1), \label {Ireg}
\end{eqnarray}
where $\zeta _R$ is the Riemann zeta function (see \cite{GrRy}).

By substituting the first term into (\ref{T00cc}), we can see that the divergent part of $\langle \hat{T}_{00}\rangle $ is proportional to $e^{-2\sigma} \Lambda$. This can be absorbed into the $(0,0)$ component of the Einstein tensor (\ref{einstein}), and is the reason for including $y$-dependence in the cutoff. Intuitively, this comes from the energy/momentum scaling due to the warping in the extra dimension. In making the $D$-dimensional Fourier transform (\ref{greenfn}), we have singled out the extra dimension. The coordinate $y$ measures coordinate length in this dimension, rather than proper length, the difference involving the warp factor. Something similar was performed in \cite{GoldbergerRothstein}, in which a rescaled cutoff was used on the $y=\pi r$ brane, from a $D$-dimensional point of view.

After these considerations, we can drop the divergent term and obtain
\begin{equation}
  \langle \hat{T}_{00}\rangle _{ren} = -\frac{1}{2}\pi ^{-1/2}(4\pi )^{-D/2}\left( \frac{ka}{1-a} \right) ^{D+1}e^{(D-1)\sigma }\Gamma \left( \frac{1+D}{2} \right) \zeta _R (1+D). \label{ccresult00}
\end{equation}

Similarly,
\begin{equation}
  \langle \hat{T}_{yy}\rangle _{ren} = \frac{D}{2}\pi ^{-1/2}(4\pi )^{-D/2}\left( \frac{ka}{1-a} \right) ^{D+1}e^{(D+1)\sigma }\Gamma \left( \frac{1+D}{2} \right) \zeta _R (1+D). \label{ccresultyy}
\end{equation}

This result, for the stress tensor for a massless, conformally coupled scalar field, agrees with what would be obtained by performing the calculation via a conformal transformation to flat spacetime.  Bearing in mind that $\lim _{k \rightarrow 0} \frac{ka}{1-a} = \frac{1}{\pi r}$, and setting $D=3$, we obtain the well-known scalar Casimir result for parallel plates in a $(3+1)$-dimensional flat background:
\begin{equation}
  \langle \hat{T}_{00}\rangle _{ren} ^{D=3, k=0} = - \frac{\pi ^2}{1440\, \ell ^4}, \label{cas}
\end{equation}
where we have put $\ell = \pi r$.

\section{General mass and coupling}

As mentioned above, the calculation can in general only be carried out numerically. To begin with, we choose particular values of interest for the dimension, mass and $\xi$, and then obtain the forms of the Green function (\ref{FTgreenfn}) and its derivatives for those values. These functions are then substituted into the integral expression for the stress tensor (\ref{genT}). We also set $k=1$, so that we work in units of the warp scale, which is of the order of the Planck scale.

As before, the integral (\ref{genT}) for $\langle \hat{T}_{\hat{\mu}\hat{\nu}} \rangle$ is ultra-violet divergent. In order to renormalise the integral, the asymptotic behaviour of the integrand for large $p$ needs to be found.

First, we make the replacements \cite{abrsteg}
\begin{eqnarray}
  J_{\nu}(iz) & = & i^\nu I_{\nu}(z), \nonumber \\
  H_{\nu}^1 (iz) & = & \frac{2}{\pi i^{\nu +1}}K_{\nu}(z), \label{IandK}
\end{eqnarray}
where $I_{\nu}$ and $K_{\nu}$ are the modified Bessel functions, so that we can use the following for large $z$ \cite{flachitoms}
\begin{eqnarray}
  I_{\nu}(z) & = & \frac{e^z}{\sqrt{2\pi z}}\Sigma ^I (z), \nonumber \\
  K_{\nu}(z) & = & \sqrt{\frac{\pi}{2z}}e^{-z}\Sigma ^K (z), \label{sigmas}
\end{eqnarray}
with
\begin{equation}
\Sigma^I (z)\simeq\sum_{k=0}^{\infty}\alpha_k\,z^{-k}\;,\label{approxSigI}
\end{equation}
where
\begin{equation}
\alpha_k=\frac{(-1)^k\,\Gamma(\nu+k+\frac{1}{2})}{2^k\,k!\,\Gamma(\nu-k+\frac{1}{2})},
\end{equation}
and
\begin{equation}  
\Sigma^K (z)\simeq\Sigma^I (-z).
\end{equation}
In practice, we take the first six terms in the sum (\ref{approxSigI}).

At this stage, we choose a value for $a$, and use the {\it Maple VI} {\tt asympt} function to find the large $p$ behaviour. In this way, we obtain the asymptotic behaviour of the integrand in (\ref{genT}), for which we write $t(p)$, as a power series:
\begin{equation}
  t(p)\sim \alpha p+d_0 +\frac{d_1}{p}+\frac{d_2}{p^2}+\frac{d_3}{p^3}+\frac{d_4}{p^4}+... \label{asympt}
\end{equation}
Now, we have to calculate an integral of the form
\begin{equation}
  \int d^D p\, t(p) \sim \int _0^\infty dp\, p^{D-1}t(p).
\end{equation}
Therefore, in order to render the integral finite, we need to subtract off terms of the asymptotic series (\ref{asympt}) up to and including the term of order $1/p^D$.

However, this introduces a problem: the final term in the subtraction diverges for {\it small} $p$. This is remedied by multiplying this term by $e^{-\mu /p}$, where $\mu$ will be the renormalisation scale.

The value of the integral is then dependent upon $\mu$. We now add and subtract a term $(d_D /p^D ) e^{-1/p}$. This allows us to isolate the renormalisation scale dependence:
\begin{eqnarray}
  \left[ \int _0^\infty dp\, p^{D-1}t(p) \right] _{ren} & = & \int _0^\infty dp\, p^{D-1} \left[ t(p) - \alpha p-d_0 \right. \nonumber \\
  &   & \left. -\frac{d_1}{p}-...-\frac{d_D}{p^D}e^{-1/p}\right] + d_D \ln \mu , \label{renormint}
\end{eqnarray}
where we have used
\begin{equation}
  \int _0^\infty dp\, \frac{1}{p}(e^{-1/p}-e^{-\mu /p}) = \ln \mu .
\end{equation}
We then calculate the renormalised integral (\ref{renormint}) numerically for a set of $y$-values.

In order to test the numerical technique, we have used it to re-calculate the stress tensor in the conformal case. We have chosen $a=10^{-1}$, as although a ``realistic'' value would be $a\sim 10^{-15}$, a smaller negative order of magnitude allows the computer to calculate effectively and allows the features of the results to be seen more readily, without qualitatively changing the results. This value of $a$ places the branes at $y=0$ and $y=2.302585...$. We also take $D=4$, $m=0$ and $\xi =3/16$.

\begin{figure}[htb]
\begin{center}
\leavevmode
\epsfxsize=100mm
\epsffile{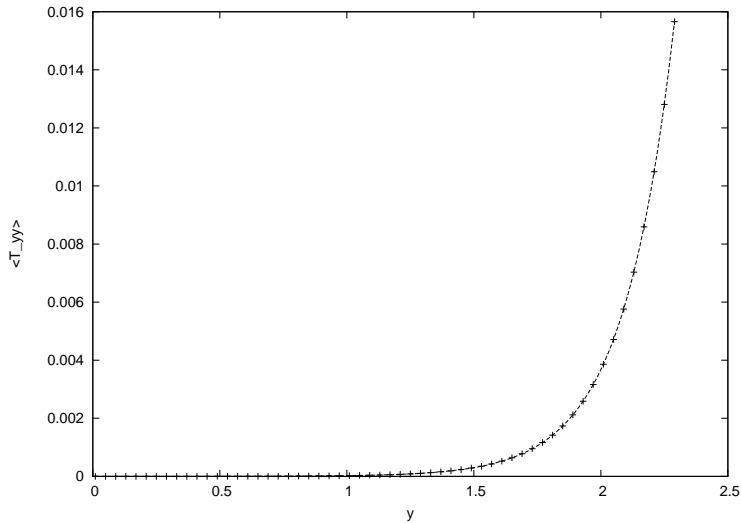}
\end{center}
\caption{\footnotesize The conformally-coupled case in (4+1) dimensions. The perfect agreement between the exact expression for $\langle \hat{T}_{yy}\rangle _{ren}$ (dashed line) and the numerical data (points) is illustrated. The branes are located at $y=0$ and $y\approx 2.3$. $k$ is set to 1 and $a=0.1$.}\label{ccplot}
\end{figure}

The results are plotted in Figure \ref{ccplot}, together with the exact analytical expression (\ref{ccresultyy}). It can be seen that the numerical results agree perfectly with the exact calculation.

Having established that the method works, we move to the massless, minimally-coupled case. This is the next simplest case, though already the calculation cannot be performed exactly. We take the same parameters as above for $a, D$ and $m$, though we now set $\xi =0$.

The results are plotted in Figure \ref{mcplot}. It is evident that the stress tensor diverges near the branes. We have also examined other cases with various values of $m$ and $\xi$, obtaining similar results.

\subsection{Surface divergence}

In \cite{deutschcandelas}, Deutsch and Candelas have argued that surface divergences in the Casimir effect arise from the high wavenumber (momentum) behaviour of the Fourier representation of $\langle \hat{T}_{\hat{\mu}\hat{\nu}} \rangle$. Taking this hint, we use it to help us isolate the part of the integrand $t(p)$ responsible for our brane divergence.

We again take our expression for $t(p)$ in terms of the functions (\ref{sigmas}). This time, we only take the first term in the sum (\ref{approxSigI}). We obtain a sum of terms of the form $p^n\exp (f(p,y))$, where $n$ is some power and $f(p,y)$ is some function in which $p$ is linear. We pick out the terms with the highest power $n$. Then, we notice that $f(p,y)$ falls into two categories concerning its $y$-dependence: either it is linear in $y$ or contains $e^{\sigma(y)}$.

We need to consider what happens to each of these kinds of term when integrated over momentum space. The first simply give either terms divergent for all $y$, part of what is removed by renormalisation, or give terms finite for all $y$. The second kind of term, however, is more complicated. These terms will turn out to be responsible for the brane divergence. We therefore discard the terms of the first kind.

After collecting the remaining terms and simplifying, it turns out that we are left with just two terms:
\begin{equation}
  t(p)_{div} \sim e^{(D-1)\sigma}p\left[ e^{2p(e^{\sigma} -1/a)}+e^{2p(1-e^{\sigma})}\right] . \label{brndiv}
\end{equation}
The first term is responsible for the $y=\pi r$ brane divergence and the second for the $y=0$ brane divergence. Integrating over $p$ space, and taking the limit as $y$ approaches the appropriate brane, we find the following expressions for the near-brane behaviour:
\begin{eqnarray}
  \langle \hat{T}_{00}\rangle _{ren}|_{y\rightarrow 0} & = & \frac{\alpha}{y^{D+1}}, \nonumber \\
  \langle \hat{T}_{00}\rangle _{ren}|_{y\rightarrow \pi r} & = & \frac{\alpha\, a^2}{(\pi r-y)^{D+1}}, \label{brndiv2}
\end{eqnarray}
where
\begin{equation}
  \alpha = (-1)^{D+1}2^{-(D+2)}\pi^{-(D+1)/2}\, (D-4\xi D-1)\; \Gamma \left( \frac{D+1}{2}\right) .
\end{equation}

\begin{figure}[htb]
\begin{center}
\leavevmode
\epsfxsize=100mm
\epsffile{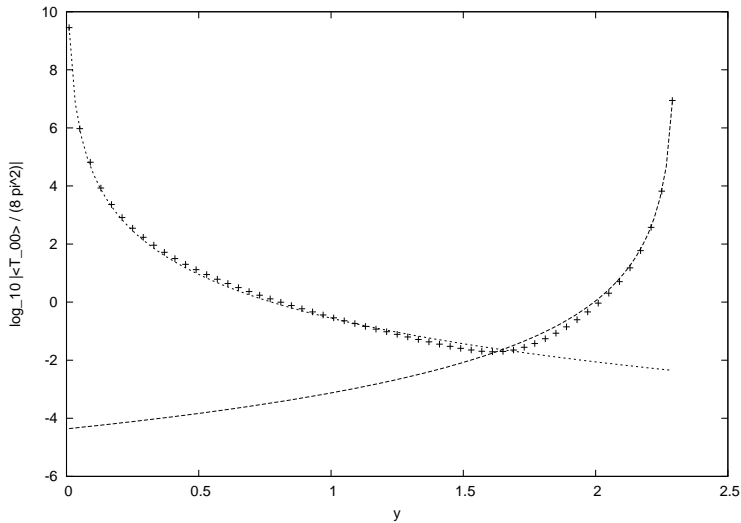}
\end{center}
\caption{\footnotesize The (4+1)-dimensional massless, minimally-coupled case. The points show the data for $\langle \hat{T}_{00}\rangle _{ren}$ from the numerical calculation. The dashed lines represent the analytic expressions for the leading near-brane behaviour. The parameters $a$ and $k$ and the positions of the branes are the same as for Figure \ref{ccplot}.}\label{mcplot}
\end{figure}

\begin{figure}[htb]
\begin{center}
\leavevmode
\epsfxsize=100mm
\epsffile{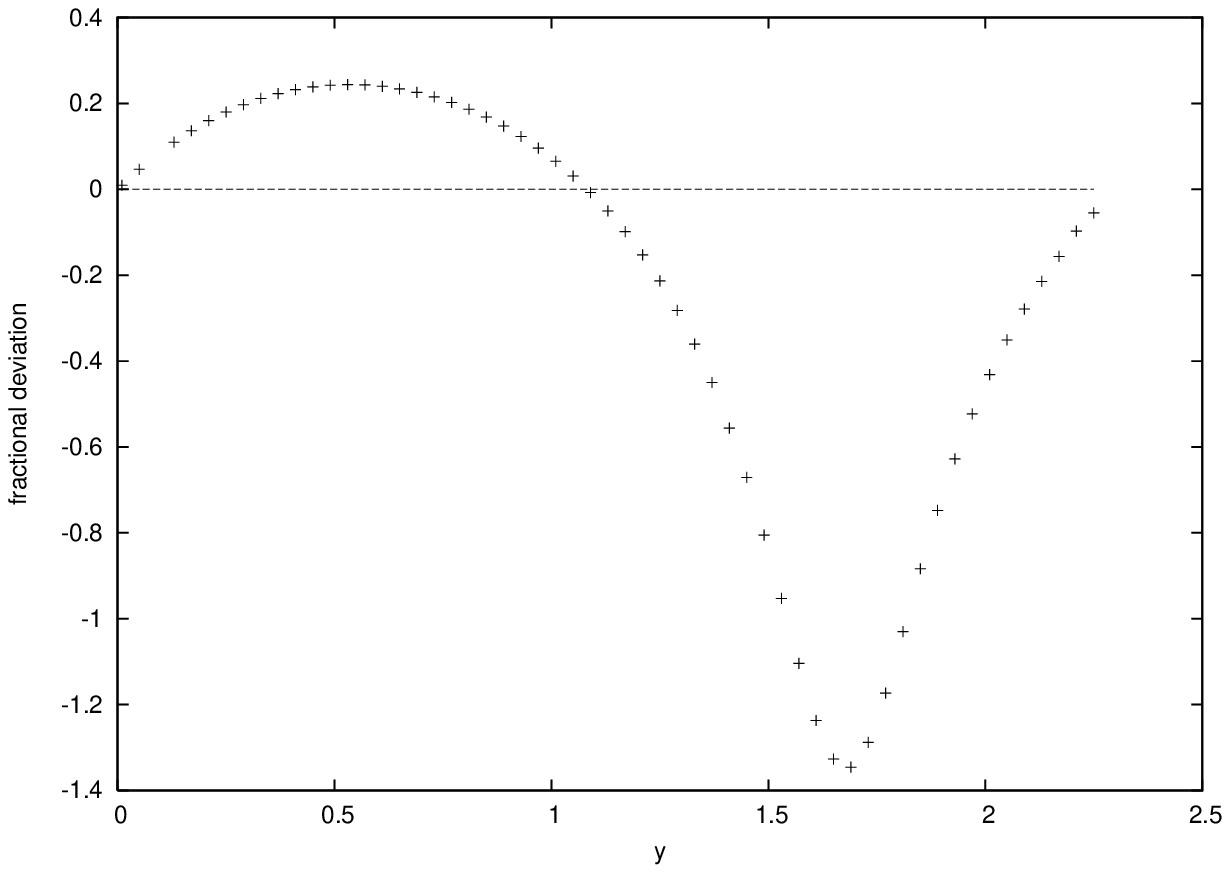}
\end{center}
\caption{\footnotesize The deviations from the near-brane behaviour. The difference between the near-brane curves and data points in Figure \ref{mcplot} are plotted as a fraction of the data, showing the convergence as the branes are approached.}\label{devplot}
\end{figure}

The dashed lines in Figure \ref{mcplot} represent these expressions in the massless, minimally coupled case, which fit the numerical results well close to the branes. Figure  \ref{devplot} shows the fractional deviations of the numerical data from the near-brane expressions.

It should be noted that the divergences {\it only} disappear for $\xi =\frac{D-1}{4D}$, i.e. in the conformally coupled case, consistently with what we found earlier.

At face value, our result that the bulk stress-energy tensor diverges non-integrably as one approaches the branes would seem to contradict the finite result for the total vacuum energy as calculated in \cite{flachitoms}, in which the renormalised sum of the zero point energies for each mode of the bulk field was calculated.

Similar problems regarding the Casimir effect between conducting surfaces in flat space were first analysed in detail by Deutsch and Candelas \cite{deutschcandelas}. They argued that this mode-sum energy is generically different from the integral of the renormalised energy density $\langle T_{00} \rangle _{ren}$, because the operation of forming the volume integral does not in general commute with that of renormalisation. Following this analysis, Kennedy, Critchley and Dowker \cite{kennedy} showed that the resolution lies in renormalisation of surface (on-boundary) terms in the effective action, which does not allow a local interpretation. This work has recently been extended to generally-coupled scalar fields for general bulk and boundary geometries \cite{saharian}.

Essentially, our stress-energy tensor is strictly-speaking incomplete: the full renormalised tensor is identical to ours, but contains an infinite term {\em on} the boundary. On integration, this surface term exactly balances the infinite contribution from the local divergences in the bulk part of the stress-energy tensor. The result is then equal to the sum of the zero-point energies.

\section{Discussion and conclusions}

In this paper, we have calculated the stress-energy tensor for a quantised bulk scalar field coupled non-minimally to the curvature in the background spacetime of the RS model. While an analytic result is not possible in general, the case of a massless, conformally-coupled field is calculated exactly. In the general case, we have produced numerical results that reveal the stress-energy to diverge near the branes, and have obtained analytic expressions for the leading behaviour of these divergences.

\subsection*{Self-Consistency}

Self-consistency requires that the RS metric (\ref{rmetric}) be a solution of the quantum corrected Einstein equations. This means including the bulk field stress-energy tensor on the right hand side of the Einstein equations with the RS metric as an ansatz (with general $\sigma (y)$) and solving for $\sigma (y)$. For the original RS metric to be strictly self-consistent, the result for $\sigma (y)$ should be of the form $k|y|$. The constant $k$ might not take its classical value, but may include a small quantum correction, this correction being the same for every component of the Einstein equations.

In the case of conformal coupling, it can be observed immediately that the traceless stress-energy tensor has a different form than the Einstein tensor, and hence each component will give a different correction to $k$, if indeed a RS-type solution can be found. This in itself implies a lack of strict self-consistency. However, if the stress-energy tensor is small (this is in fact the case: e.g.~$|\langle \hat{T}_{00} \rangle _{ren}| \sim 10^{-30}k^5$ at most for $D=4$), then we might expect the RS solution to be {\em approximately} self-consistent, in the sense that $\sigma(y)=k|y|$ plus extra small terms due to quantum effects, which may differ for each component. The true self-consistent solution might then not differ from the original RS model significantly.

In the general case, the stress-energy tensor diverges close to the branes. This would also be expected in the conformal case if the branes are curved, however slightly \cite{deutschcandelas}. Although the generic stress-energy tensor is not traceless, it would also not be expected to have to same form as the Einstein tensor, except possibly in some special case. Therefore, any argument relying upon smallness of the stress-energy tensor will not apply, and a conclusion is then that, generically, the RS metric is not self-consistent, even approximately, in the presence of a bulk scalar field. If there is a self-consistent solution, then $\sigma (y)$ will not approximate to $k|y|$.

As mentioned above, the Casimir effect also suffers from divergences. However, the imperfect nature of the conducting surfaces becomes apparent (in the form of a finite skin-depth \cite{deutschcandelas}) and destroys the divergence long before the backreaction of the fields between the plates becomes measurable. In our case, the analogue is the quantum ``imperfection'' of the branes themselves, on the order of the Planck scale, at which quantum-gravity effects would come into play and alter the situation. Due to the $1/y^{D+1}$ behaviour of the divergences, the backreaction should become significant before this region is reached.

It should be noted that this conclusion might not hold in a supersymmetric theory, where a fermionic contribution could exactly cancel the bulk field vacuum energies. We have also not taken into account the effect of the surface terms in the stress-energy tensor. Furthermore, it would be interesting to examine the consequences of introducing a potential to the scalar field action.

Whether or not the RS model with bulk fields can be rendered self-consistent, the main conclusion of this paper is that the stress-energy tensor is sufficiently complicated that it has implications for the effective potential methods that have so far been used. The important difference with the older Kaluza-Klein scenario of Candelas and Weinberg \cite{candelasweinberg} is that assuming spheres for the extra dimensions, as in that case, results in a homogeneous space for which the stress tensor is determined by symmetry from the vacuum energy. For the RS model, homogeneity is lacking so that it is not necessarily sufficient to use an effective potential in general.

\section*{Acknowledgements}

A.~Knapman is grateful to W.~Naylor and A.~Flachi for useful comments and to I.~G.~Moss for helpful discussions, particularly on the stress-energy divergences.

\end{document}